\documentclass[11pt]{article}
\usepackage{amsmath, amsthm, amsfonts}
\usepackage[margin=1in,footskip=0.25in]{geometry}
\usepackage{makecell}
\usepackage{multicol}
\usepackage{graphicx}
\theoremstyle{theorem}

\newtheoremstyle{defi}
  {10pt}          
  {10pt}  
  {\rm}  
  {\parindent}     
  {\bf}  
  {. }    
  { }    
  {}     
\theoremstyle{defi}

\begin{document}

\date{}

\title{\bf Kaluza-Klein bulk viscous fluid cosmological models and the validity of the second law of thermodynamics in $f(R, T)$ gravity}
\author{G. C. Samanta$^{1}$, R. Myrzakulov$^{2}$ and Parth Shah$^{3}$\\
 $^{1, 3}$Department of Mathematics\\
BITS Pilani K K Birla Goa Campus,\\
Goa-403726, India,\\ gauranga81@gmail.com\\
$^{2}$Eurasian International Center for Theoretical Physics\\ and Department of General Theoretical Physics,\\ Eurasian National University, Astana 010008, Kazakhstan\\
rmyrzakulov@gmail.com}

\maketitle

\begin{abstract} The authors considered the bulk viscous fluid in $f(R, T)$ gravity within the framework of Kaluza-Klein space time. The bulk viscous coefficient $(\xi)$ expressed as  $\xi=\xi_0+\xi_1\frac{\dot{a}}{a}+\xi_2\frac{\ddot{a}}{\dot{a}}$, where $\xi_0$, $\xi_1$ and $\xi_2$ are positive constants. We take $p=(\gamma-1)\rho$, where $0\le\gamma\le2$ as an equation of state for perfect fluid. The exact solutions to the corresponding field equations are given by assuming a particular model of the form of $f(R, T)=R+2f(T)$, where $f(T)=\lambda T$, $\lambda$ is constant. We studied the cosmological model in two stages, in first stage: we studied the model with no viscosity, and in second stage: we studied the model involve with viscosity. The cosmological model involve with viscosity is studied by five possible scenarios for bulk viscous fluid coefficient $(\xi)$.  The total bulk viscous coefficient seems to be negative, when the bulk viscous coefficient is proportional to $\xi_2\frac{\ddot{a}}{\dot{a}}$, hence the second law of thermodynamics is not valid, however, it is valid with the generalized second law of thermodynamics. The total bulk viscous coefficient seems to be positive, when, the bulk viscous coefficient is proportional to $\xi=\xi_1\frac{\dot{a}}{a}$, $\xi=\xi_1\frac{\dot{a}}{a}+\xi_2\frac{\ddot{a}}{\dot{a}}$ and $\xi=\xi_0+\xi_1\frac{\dot{a}}{a}+\xi_2\frac{\ddot{a}}{\dot{a}}$, so the second law of thermodynamics and the generalized second law of thermodynamics is satisfied throughout the evolution. We calculate statefinder parameters of the model and observed that, it is different from the $\wedge$CDM model. Finally, some physical and geometrical properties of the models are discussed.
\end{abstract}



\section{Introduction}
The idea of the compactification of extra space dimensions of space-time, is known as Kaluza-Klein theories (\cite{kaluza}, \cite{klein}). The interesting idea of Kaluza-Klein theory is to unify the gravity and other interactions. Let us define the Kaluza-Klein action in $d=(1+(d-1))$ dimensions which will describe the massless fermionic field and only pure gravity. Such an action is defined as
\begin{equation}\label{1}
  S_{KK}=S_D+S_E=\int d^dxE(\frac{1}{2}\Psi^{\dagger}\gamma^0\gamma^ap_{0a}\Psi+h.c.)-\alpha\int d^dxER,
\end{equation}
where $h.c.$ means Hermitian conjugate, $S_E$ is the Einstein action describing gravity, $E=det(-g_{\mu\nu})^{\frac{1}{2}}$, $\alpha$ is the gravitational coupling constant and $R$ is the Ricci scalar.
\par
The original Kaluza-Klein theory derive with one extra spatial dimension. The appropriate metric tensor for five dimensional space time is
\begin{equation}\label{2}
  \hat{g}_{\mu\nu}=\left(\begin{array}{cc}
                           g_{\mu\nu} & g_{\mu5}  \\
                           g_{\nu5} & g_{55}
                         \end{array}\right).
\end{equation}
The fifth dimension is postulated to be comfactified, rolled-up in a small circle, which provides us the explanation for the un-observability of the extra dimension. Hence the topology of the five dimensional space time is $M^4\times S^1$, where $M^4$ is the standard four-dimensional Minkowski space-time and $S^1$ is a circle with very small radius. The simplest way to imagine space with one extra dimension is to imagine a small circle at every point of 3-dimensional space.
\par
Inflation is an important idea in cosmology. There are two scenarios proposed in Kaluza-Klein cosmology. The first scenario \cite{shafi} is: the scale of standard 3-dimensional space expands when the scale of internal space changes slowly with time. The second scenario (\cite{sahdev}, \cite{abbott}) is: inflation occurs near the singularity $a(t)\rightarrow\infty,~~b(t)\rightarrow 0 ~~(t\rightarrow t_0)$.
\par
Expansion of our universe is in an accelerating way which is suggested by type Ia supernova observational data (\cite{riess}, \cite{perlmutter}). Myrzakulov \cite{myrzakulov3} constructed several concrete models describing the trefoil and figure-eight knot universes from Bianchi-type I cosmology and examined the cosmological features and properties in detail. Yesmakhanova et al \cite{Yesmakhanova} constructed a cosmological model by assuming the periodic forms for pressure and energy density as a functions of time, there exists a coordinate set, in which the time evolutions of the space is knot like.
Very recent, the concept of viscosity is introduced into dark energy study. Now, it seems to play a more and more important role in constructions of cosmological model. The concept of viscosity has come from fluid mechanics, it is related to the velocity gradient of the fluid and is divided into two classes, bulk viscosity and shear viscosity. Shear viscosity is related to anisotropic space-time. Bulk viscosity usually related to isotropic space time.
 Misner \cite{misner} pointed out that during cosmic evolution when neutrinos decouple from the cosmic fluid bulk viscosity could arise and lead to an effective mechanism of entropy production. The isotropic homogeneous spatially flat cosmological
model with bulk viscous fluid discussed by \cite{murphy}.
 Bulk viscosity related to the grand unified-theory phase transition \cite{Langacher} may lead to explain the cosmic acceleration. The presence of bulk viscosity leads to an inflationary like solutions in FRW space-time obtained by \cite{Padmanabhan 1987}. Johri and Sudarshan \cite{Johri} studied that the presence of bulk viscosity leads to an inflationary universe in Brans-Dicke theory. Brevik et al. \cite{Brevik} prove, in particular, that a viscous fluid 
 is perfectly able to produce a Little Rip cosmology as a purely viscosity effect. Myrzakul et al\cite{Myrzakul} studied inhomogeneous viscous fluids cosmological model in flat FRW space time, in which authors discussed the presence of finite future time singularities. Subsequently, the cosmological models involve with viscous fluid are constructed and discussed by many authors (\cite{Mohanty1}, \cite{Mohanty2}, \cite{Pavon1}, \cite{Pavon2}, \cite{Burd}, \cite{Marteens}, \cite{Zimdahl1}, \cite{Zimdahl2}, \cite{Chimento1}, \cite{Chimento2}, \cite{Pradhan 2001}, \cite{Bali}, \cite{Wang1}, \cite{Wang2}, \cite{Fabris}, \cite{Saha}, \cite{Tripathy}, \cite{Katore2011}, \cite{Saadat}, \cite{sasidharan}, \cite{mahanta2014}
   ). Recently Avelino and Nucamendi \cite{avelino} explained the present cosmic acceleration of the universe through bulk viscous fluid by taking the constant bulk viscous coefficient. In this work, the model proposed by Avelino and Nucamendi \cite{avelino} has been extended and improved upon to reflect the more general situation. We extend their work into Kaluza-Kelin space-time and the coefficient of bulk viscous fluid is proportional to the linear combination of three terms, such as $\xi=\xi_0+\xi_1\frac{\dot{a}}{a}+\xi_2\frac{\ddot{a}}{\dot{a}}$, (where $\xi_0$, $\xi_1$ and $\xi_2$ are constants) rather than $\xi=\xi_0+\xi_1\frac{\dot{a}}{a}$.

\section{Review of $f(R, T)$ gravity}
The $f(R, T)$ gravity is a modification of Einstein gravity, in which the Einstein-Hilbert Lagrangian, i. e. $R$ is replaced by
an arbitrary function $R$ and $T$, where $R$ is the curvature scalar and $T$ is the trace of energy momentum tensor. The modification of Einstein theory is proposed by \cite{harko}. \\

\begin{equation}\label{3}
  S=\frac{1}{16 \pi}\int\left(f(R, T)+16\pi L_m\right)\sqrt{-g}d^4x
\end{equation}
is the action for the $f(R, T)$ modified gravity, where $g$ is the determinate of the metric tensor $g_{\mu\nu}$ and $L_m$ is the matter Lagrangian density. \\
Here, we consider the system of units where $G=c=1$. The field equations for the $f(R, T)$ modified gravity are obtained from the action $S$ given in equation \eqref{4} as
\begin{eqnarray} \label{4}
  f_R(R, T)(R_{\mu\nu}-\frac{1}{3}Rg_{\mu\nu})+\frac{1}{6}f(R, T)g_{\mu\nu}&=&
  8\pi (T_{\mu\nu}-\frac{1}{3}Tg_{\mu\nu})-f_T(R, T)(T_{\mu\nu}-\frac{1}{3}Tg_{\mu\nu})\nonumber \\
  &-& f_T(R, T)(\Theta_{\mu\nu}-\frac{1}{3}\Theta g_{\mu\nu})+\bigtriangledown_{\mu}\bigtriangledown_{\nu}f_R(R, T),
\end{eqnarray}
where $\Theta=g^{\mu\nu}\Theta_{\mu\nu}$, $\Box \equiv \bigtriangledown_{\mu} \bigtriangledown^{\mu}$ is the d'Alembert operator, $\bigtriangledown_{\mu}$ being the covariant derivative. $f_R(R, T)=\frac{\partial f(R, T)}{\partial R}$ and $f_T(R, T)= \frac{\partial f(R, T)}{\partial T}$ are partial derivative with respect to $R$ and $T$ respectively.
If we consider the matter of the universe as a perfect fluid, then the stress energy momentum tensor of the matter Lagrangian is obtained by
\begin{equation}\label{5}
  T_{\mu\nu}=(p+\rho)u_{\mu}u_{\nu}-pg_{\mu\nu},
\end{equation}
and the matter Lagrangian can be taken as $L_m=-p$. The four velocity vector in co-moving co-ordinates system is defined as
$u^{\mu}=(1, 0, 0, 0)$ which satisfies the conditions $u_{\mu}u^{\mu}=1$ and $u^{\mu} \bigtriangledown_{\nu}u_{\mu}=0$.
Here $p$ and $\rho$ are the isotropic pressure and energy density of the universe respectively. We find the stress-energy of a perfect fluid as
\begin{equation}\label{6}
  \Theta_{\mu\nu}=-2T_{\mu\nu}-pg_{\mu\nu}.
\end{equation}
Now, we can find the different theoretical models for the different choice of $f(R, T)$. Harko et al.\cite{harko} considered three different explicit form of $f(R, T)$ as
\begin{equation}\label{7}
  f(R, T)=\begin{cases}
            R+2f(T),  \\
             f_1(R)+f_2(T), \\
            f_1(R)+f_2(R)f_3(T).
                      \end{cases}
\end{equation}
Myrzakulov \cite{myrzakulov2012} presented a new method to construct particular models of $f(R, T)$ gravity and considered the $M_{43}$-model, deriving its action in terms of the curvature and torsion scalars. Then in detail author studied the $M_{37}$-model and presented its action, Lagrangian and equations of motion for the FRW space time. Finally, they shown that the model can describes the accelerated expansion of the Universe. Myrzakulov \cite{myrzakulov1} studied dark energy in  $f(R, T)$ gravity and showed that for some values of the parameters the expansion of the universe can be
accelerated without introducing any dark component.
Subsequently many authors (\cite{mahanta2014}, \cite{singh3}, \cite{adhav2012}, \cite{houndjo2012}, \cite{mishra2014}, \cite{Reddy2012}, \cite{samanta2013}, \cite{samanta2013a}, \cite{samanta2013b},  \cite{singh2014}, \cite{ahmed}, \cite{pradhan}, \cite{katore}, \cite{Rao1}, \cite{Rao2}, \cite{Rao3}, \cite{Rao4}, \cite{das}, \cite{zubair}, \cite{singh1}, \cite{singh2}, \cite{Momeni}, \cite{JamilM}, \cite{JamilM1}, \cite{MomeniD}, \cite{Gu}, \cite{Yousaf}, \cite{Yousaf1}) studied some cosmological models in $f(R, T)$ modified gravity for different choice of $f(R, T)$ from various angles.
\par
In this paper, authors consider the following form of $f(R, T)$
\begin{equation}\label{8}
  f(R, T)=R+2f(T).
\end{equation}
 The term $2f(T)$ in the gravitational action modifies the
gravitational interaction between matter and curvature scalar $R$. Using equation \eqref{8}, one can re-write the gravitational field
equations defined in \eqref{4} as
\begin{equation}\label{9}
  R_{\mu\nu}-\frac{1}{2}R g_{\mu\nu}=8\pi T_{\mu\nu}-2f^{'}(T)(T_{\mu\nu}+\Theta_{\mu\nu})+f(T)g_{\mu\nu},
\end{equation}
which is considered as the field equation of $f(R, T)$ gravity for the above particular form of $f(R, T)$. Here the prime stands for
derivative of $f(T)$ with respect to $T$. This is the $f(R, T)$ gravity field equations for the particular case $f(R, T)=R+2f(T)$.
\section{General formalism of equations of motion}
  Here we consider the Kaluza-Klein type space-time metric in the form
  \begin{equation}\label{10}
    ds^2=dt^2-e^c(dr^2+r^2d\theta^2+r^2\sin^2\theta d\phi^2)-e^bd\Psi^2,
  \end{equation}
  where $c(t)$ and $b(t)$ are time dependent cosmological scale factors. \\
  The mean scale factor $(a)$ of the cosmological model \eqref{10} is defined as
  \begin{equation}\label{11}
    a=\left(e^{\frac{3c+b}{2}}\right)^{\frac{1}{4}}.
  \end{equation}
  The spatial volume $(V)$ for the space-time \eqref{10} is defined as
  \begin{equation}\label{12}
    V=e^{\frac{3c+b}{2}}.
  \end{equation}
  The mean Hubble parameter $(H)$ for the space-time \eqref{10} is defined as
  \begin{equation}\label{13}
    H=\frac{1}{8}(3\dot{c}+\dot{b}).
  \end{equation}
  \par
  In this work, we consider the source of gravitation is a combination of perfect fluid and bulk viscous
fluid. Therefore, the energy momentum tensor takes the form
\begin{equation}\label{14}
  T_{\mu\nu}=(\rho+\bar{p})u_{\mu}u_{\nu}-\bar{p}g_{\mu\nu}
\end{equation}
and
\begin{equation}\label{15}
  \bar{p}=p-4\xi H,
\end{equation}
where $\rho$ is the energy density, $\xi$ is the coefficient of bulk viscosity, $\bar{p}$ is effective pressure, $p$ is the proper pressure and $u^{\mu}=(1, 0, 0, 0, 0)$ being the five velocity vector of the fluid on the co-moving coordinates. From the
thermodynamical point of view, $\xi$ is chosen to be positive and may depend on time $t$, or the scale factor, or the energy density $\rho$, etc.
Here $H=\frac{\dot{a}}{a}$ is Hubble parameter, where an over dot stands for the derivative with respect to time 't'. Hence, the Lagrangian density
may be chosen as $L_m=-\bar{p}$ and the tensor $\Theta_{\mu\nu}$ in \eqref{6} reduces to
\begin{equation}\label{16}
  \Theta_{\mu\nu}=-2T_{\mu\nu}-\bar{p}g_{\mu\nu}.
\end{equation}
Now using the equations \eqref{14} and \eqref{16}, the field equation \eqref{9} for bulk viscous fluid is given by
\begin{equation}\label{17}
  R_{\mu\nu}-\frac{1}{2}Rg_{\mu\nu}=8\pi T_{\mu\nu}+2f^{'}(T)T_{\mu\nu}+(2\bar{p}f^{'}(T)+f(T))g_{\mu\nu}.
\end{equation}
Using co-moving coordinates and equation \eqref{14}, the gravitational field equations of $f(R, T)$ gravity \eqref{17} with the particular choice of the function $f(T)=\lambda T$, where $\lambda$ is an arbitrary constant, for the Kaluza-Klein type space-time \eqref{10} are given by
\begin{equation}\label{18}
  -\frac{3}{4}(\dot{c}^2+\dot{c}\dot{b})=(8\pi+3\lambda)\rho-2\bar{p}\lambda,
\end{equation}
\begin{equation}\label{19}
  \ddot{c}+\frac{3}{4}\dot{c}^2+\frac{1}{2}\ddot{b}+\frac{1}{4}\dot{b}^2+\frac{1}{2}\dot{c}\dot{b}=(8\pi+4\lambda)\bar{p}-\lambda\rho
\end{equation}
and
\begin{equation}\label{20}
  \frac{3}{2}(\ddot{c}+\dot{c}^2)=(8\pi+4\lambda)\bar{p}-\lambda\rho.
\end{equation}
The equation of continuity is given by
\begin{equation}\label{21}
  \dot{\rho}+4H(\bar{p}+\rho)=0.
\end{equation}
Through the comparisons of viscous and non-viscous models, it is beneficial for us to understand the role of cosmic viscosity, the properties of the cosmic models with common EoS and further our physical universe more comprehensively.
\section{Cosmology solution with perfect fluid i. e. no viscosity}
In this section, we discuss the model with non viscous fluid. We can choose the equation of state in the following form
\begin{equation}\label{22}
  p=(\gamma-1)\rho,
\end{equation}
where $\gamma$ is constant known as the EoS parameter lying in the range $0\le\gamma\le2$. In order to get an exact solution of the field equations we use the following relation
\begin{equation}\label{23}
  c=nb,
\end{equation}
where $n$ is a non zero real constant. \\
Subtract equation \eqref{20} from equation \eqref{19} and use the condition \eqref{23}, we obtain
\begin{equation}\label{24}
  \ddot{b}+\left(\frac{3n+1}{2}\right)\dot{b}^2=0,
\end{equation}
where $n\in R-\{0, -1/3\}$. Solving equation \eqref{24}, we obtain $b=\ln\bigg[\left(\frac{3n+1}{2}\right)t-k\bigg]^{\frac{2}{3n+1}}+k_1$, where $k$ and $k_1$ are constants of integration.
Without loss of generality and for simplicity, we can choose $k_1=0$. Thus
\begin{equation}\label{25}
  b=\ln\bigg[\left(\frac{3n+1}{2}\right)t-k\bigg]^{\frac{2}{3n+1}}.
\end{equation}
Using equation \eqref{25} in \eqref{23}, we get
\begin{equation}\label{26}
  c=\ln\bigg[\left(\frac{3n+1}{2}\right)t-k\bigg]^{\frac{2n}{3n+1}}.
\end{equation}
The mean Hubble parameter is obtained as
\begin{equation}\label{27}
  H=\left(\frac{3n+1}{4}\right)\frac{1}{(3n+1)t-2k}.
\end{equation}
The physical behavior or dynamics of the standard spatial dimension $(c)$, the extra dimension $(b)$ and the Hubble parameter $(H)$ with respect to time is given in Figure-1, 2 \& 3 respectively. Please see the Figure-1, 2 \& 3.\\
Using equations \eqref{22} and \eqref{27} in \eqref{21}, we obtain $\rho=k_2[(3n+1)t-2k]^{-\gamma}$, where $k_2$ is an integration constant. Without loss of generality, we can take $k_2=1$. Thus
\begin{equation}\label{28}
  \rho=[(3n+1)t-2k]^{-\gamma}.
\end{equation}
The variation of the energy density with respect to time is given in Figure-4, we analyze the behavior of the energy density for three different types of models, say $\gamma=\frac{2}{3}$, $\gamma=1$ and $\gamma=\frac{4}{3}$, and we observed that the model $\gamma=\frac{2}{3}$ is dominated by the model $\gamma=1$ is dominated by $\gamma=\frac{4}{3}$. \\
The deceleration parameter $q$ is calculated as
\begin{equation}\label{29}
  q=\frac{-1}{2}.
\end{equation}
Generally, the statefinder parameters pair $\{r, s\}$ is given by \cite{sahni}
\begin{equation}\label{30}
  r=\frac{\dddot{a}}{aH^3},~~~~s=\frac{r-1}{3\left(q-\frac{1}{2}\right)},
\end{equation}
using the equations \eqref{11} and \eqref{27}, the value of the statefinder pair obtained as $\{r, s\}=\{\frac{21}{64}, \frac{43}{192}\}$, where $r<1$ and $s<1$, which is different from the standard $\wedge$CDM model.
\section{Cosmology with constant bulk viscous coefficient $(\xi=\xi_0)$}
In this section, we discuss the cosmological model involve with perfect fluid and bulk viscous fluid, in which the coefficient of a bulk viscous fluid has taken as a simple constant i. e. $(\xi=\xi_0)$. Now using the equations \eqref{22} and \eqref{27} in \eqref{21}, we obtain
\begin{equation}\label{31}
  \dot{\rho}+\frac{\gamma(3n+1)}{(3n+1)t-2k}\rho=\xi_0\frac{(3n+1)^2}{[(3n+1)t-2k]^2}.
\end{equation}
The general solution of the equation \eqref{31} is obtained as
\begin{equation}\label{32}
  \rho=\frac{k_3}{[(3n+1)t-2k]^{\gamma}}+\xi_0\frac{(3n+1)}{(\gamma-1)[(3n+1)t-2k]},
\end{equation}
where $k_3$ is an integration constant. \\
The dynamics of the energy density with respect to time is given in Figure-5, we analyze the behavior of the energy density for five different types of models, say $\gamma=0$, $\gamma=\frac{2}{3}$, $\gamma=1.001$, $\gamma=\frac{4}{3}$ and $\gamma=2$, and we observed that the model $\gamma=\frac{2}{3}$ is dominated by the model $\gamma=1$ is dominated by $\gamma=\frac{4}{3}$, and we observe that the model posses a type-III singularity for $\gamma=1$, i. e. $p=0$, dust case. Please see the Figure-5 to analyze the behavior of the energy density with constant bulk viscous coefficient.

\section{Cosmology with non constant bulk viscous coefficient}
In this section, authors discuss the cosmological model involve with perfect fluid and bulk viscous fluid, in which the bulk viscous coefficient $(\xi)$ is proportional to $\xi_0+\xi_1\frac{\dot{a}}{a}+\xi_2\frac{\ddot{a}}{\dot{a}}$, i. e. $\xi=\xi_0+\xi_1\frac{\dot{a}}{a}+\xi_2\frac{\ddot{a}}{\dot{a}}$. Using $H=\frac{\dot{a}}{a}$, it can be written as,
\begin{equation}\label{33}
  \xi=\xi_0+\xi_1H+\xi_2\left(\frac{\dot{H}}{H}+H\right).
\end{equation}
Here we discuss the cosmological model through the following four cases:\\
Case-I $\xi=\xi_1H$\\
Case-II $\xi=\xi_2\left(\frac{\dot{H}}{H}+H\right)$\\
Case-III $\xi=\xi_1H+\xi_2\left(\frac{\dot{H}}{H}+H\right)$\\
Case-IV $\xi=\xi_0+\xi_1H+\xi_2\left(\frac{\dot{H}}{H}+H\right)$
\subsection{Case-I $(\xi=\xi_1H)$}
Now using the equations \eqref{22} and \eqref{27} in \eqref{21}, we obtain
\begin{equation}\label{34}
  \dot{\rho}+\frac{\gamma(3n+1)}{(3n+1)t-2k}\rho=\frac{\xi_1}{4}\frac{(3n+1)^3}{[(3n+1)t-2k]^3}.
\end{equation}
The general solution of the equation \eqref{34} is obtained as
\begin{equation}\label{35}
  \rho=\frac{k_4}{[(3n+1)t-2k]^{\gamma}}+\frac{\xi_1}{4}\frac{(3n+1)^2}{(\gamma-2)[(3n+1)t-2k]^2},
\end{equation}
where $k_4$ is an integration constant.\\
The dynamics of the energy density with respect to time is given in Figure-6, we analyze the behavior of the energy density for four different types of models, say $\gamma=\frac{2}{3}$, $\gamma=1$, $\gamma=\frac{4}{3}$ and $\gamma=1.99$, and we observe that the model posses a type-III singularity for $\gamma=2$, i. e. $p=\rho$, stiff fluid case. Please see the Figure-6 to analyze the behavior of the energy density, when the bulk viscous coefficient is proportional to the expansion rate of the universe, i. e. $\xi=\xi_1\frac{\dot{a}}{a}$. \\

The bulk viscous coefficient
\begin{equation}\label{36}
  \xi=\xi_1\frac{3n+1}{4}\frac{1}{(3n+1)t-2k}.
\end{equation}
The variation of the bulk viscous coefficient $(\xi=\xi_1\frac{\dot{a}}{a})$ is given in Figure-8, we observed that, the bulk viscous coefficient $(\xi)$ is always positive and decreases to zero as $t\rightarrow\infty$.
\subsection{Case-II $\xi=\xi_2\left(\frac{\dot{H}}{H}+H\right)$}
Now using the equations \eqref{22} and \eqref{27} in \eqref{21}, we obtain
\begin{equation}\label{37}
  \dot{\rho}+\frac{\gamma(3n+1)}{(3n+1)t-2k}\rho=\frac{5\xi_2}{4}\frac{(3n+1)^3}{[(3n+1)t-2k]^3}.
\end{equation}
The general solution of the equation \eqref{37} is obtained as
\begin{equation}\label{38}
  \rho=\frac{k_5}{[(3n+1)t-2k]^{\gamma}}+5\xi_2\frac{(3n+1)^2}{(\gamma-2)[(3n+1)t-2k]^2},
\end{equation}
where $k_5$ is an integration constant. \\
The bulk viscous coefficient
\begin{equation}\label{39}
  \xi=-\xi_2\frac{3}{4}\frac{3n+1}{(3n+1)t-2k}.
\end{equation}
The variation of the bulk viscous coefficient $(\xi=\xi_2\frac{\ddot{a}}{\dot{a}})$ is given in Figure-9, we observed that, the bulk viscous coefficient $(\xi)$ is always negative and decreases to zero as $t\rightarrow\infty$.

\subsection{Case-III $\xi=\xi_1H+\xi_2\left(\frac{\dot{H}}{H}+H\right)$}
Now using the equations \eqref{22} and \eqref{27} in \eqref{21}, we obtain
\begin{equation}\label{40}
  \dot{\rho}+\frac{\gamma(3n+1)}{(3n+1)t-2k}\rho=\frac{(\xi_1+5\xi_2)(3n+1)^3}{4[(3n+1)t-2k]^3}.
\end{equation}
The general solution of the equation \eqref{40} is obtained as
\begin{equation}\label{41}
  \rho=\frac{k_6}{[(3n+1)t-2k]^{\gamma}}+\frac{(\xi_1+5\xi_2)(3n+1)^2}{4(\gamma-2)[(3n+1)t-2k]^2},
\end{equation}
where $k_5$ is an integration constant.\\
The bulk viscous coefficient
\begin{equation}\label{42}
  \xi=(\xi_1-3\xi_2)\frac{3n+1}{4[(3n+1)t-2k]}.
\end{equation}
\subsection{Case-IV $\xi=\xi_0+\xi_1H+\xi_2\left(\frac{\dot{H}}{H}+H\right)$}
Now using the equations \eqref{22} and \eqref{27} in \eqref{20}, we obtain
\begin{equation}\label{43}
  \rho=\frac{1}{(8\pi+3\lambda-\gamma)}\times\bigg[\xi_0\frac{(3n+1)(8\pi+4\lambda)}{(3n+1)t-2k}+(\xi_1-3\xi_2)\frac{(3n+1)^2}{4[(3n+1)t-2k]^2}-\frac{
  12n(n+1)}{4[(3n+1)t-2k]^2}\bigg].
\end{equation}
The variation of the energy density with respect to time is given in Figure-7, we analyze the behavior of the energy density for five different types of models, say $\gamma=0$, $\gamma=\frac{2}{3}$, $\gamma=1$, $\gamma=\frac{4}{3}$ and $\gamma=2$. \\
Using the equation \eqref{27} in \eqref{33}, the bulk viscosity coefficient $(\xi)$ can be calculated as
\begin{equation}\label{44}
  \xi=\xi_0+(\xi_1-3\xi_2)\frac{3n+1}{4}\frac{1}{(3n+1)t-2k}.
\end{equation}
The variation of $\xi=\xi_0+\xi_1\frac{a}{a}+\xi_2\frac{\ddot{a}}{\dot{a}}$ is given in Figure-10, it is observed that the bulk viscous coefficient $\xi$ is always positive and decreases to $\xi_0$ as $t\rightarrow\infty$.
\section{Second law of thermodynamics}
The local entropy in FRW space time is given by \cite{weinberg} is defined as
\begin{equation}\label{45}
  T\bigtriangledown_\mu s^\mu=\xi(\bigtriangledown_\mu u^\mu)^2=9H^2\xi.
\end{equation}
We can define for the Kaluza-Klein space time as
\begin{equation}\label{46}
  T\bigtriangledown_\mu s^\mu=\xi(\bigtriangledown_\mu u^\mu)^2=16H^2\xi
\end{equation}
where $\bigtriangledown_\nu s^\nu$ is the rate of generation of entropy per unit volume and $T$ is the temperature. The second law of thermodynamics is satisfied if,
\begin{equation}\label{47}
  T\bigtriangledown_\mu s^\mu\ge 0,
\end{equation}
which implies from equation \eqref{46} that
\begin{equation}\label{48}
  \xi\ge 0,
\end{equation}
where
\begin{equation}\label{49}
  \xi=\xi_1\frac{3n+1}{4}\frac{1}{(3n+1)t-2k},
\end{equation}
\begin{equation}\label{50}
  \xi=-\xi_2\frac{3}{4}\frac{3n+1}{(3n+1)t-2k},
\end{equation}
\begin{equation}\label{51}
  \xi=(\xi_1-3\xi_2)\frac{3n+1}{4[(3n+1)t-2k]}
\end{equation}
and
\begin{equation}\label{52}
  \xi=\xi_0+(\xi_1-3\xi_2)\frac{3n+1}{4}\frac{1}{(3n+1)t-2k}
\end{equation}
are obtained from the equations \eqref{36}, \eqref{39}, \eqref{42} and \eqref{44} respectively. From the equation \eqref{49}, we found that, the
 dynamics of the bulk viscous coefficient $(\xi)$ is continuously positive throughout the cosmic time 't'. From equations \eqref{51} and \eqref{52}, we found that, the dynamics of the bulk viscous coefficient $(\xi)$ is continuously positive throughout the cosmic time 't' for $\xi_1 \ge 3\xi_2$. Hence the entropy production rate is positive throughout the evolution of the universe. Hence the second law of thermodynamics is obey throughout the evolution. But, from the equation \eqref{50}, we found that, the bulk viscous coefficient $(\xi)$ is changing continuously from negative to zero. This indicates that, the production rate of entropy is negative. Therefore, the second law of thermodynamics is violated.
The behavior of the bulk viscous coefficient is decreasing in nature. So, the bulk viscous fluid has more significance in earlier epochs of the universe than the future.
\par
The generalized second law of thermodynamics state that, the total entropy of the fluid components plus that of the horizon of the universe always increase or positive constant \cite{Gibbons}. This implies, the rate of change of entropy of the bulk viscous fluid and that of the horizon must be positive.
\begin{equation}\label{53}
  \frac{d}{dt}(S_m+S_h)\ge 0,
\end{equation}
where $S_m$ is the entropy of the matter and $S_h$ is that of the horizon. The apparent horizon radius is defined as \cite{Sheykhi}
\begin{equation}\label{54}
  r_A=\frac{a}{\dot{a}}.
\end{equation}
The entropy together with apparent horizon is defined by \cite{davis} is
\begin{equation}\label{55}
  S_h=2\pi A=8\pi^2 r^2_A,
\end{equation}
where $A=4\pi r^2_A$ is the area of the apparent horizon. Using the equations \eqref{15}, \eqref{21} and \eqref{33}, we obtain,
\begin{equation}\label{56}
  \dot{r_A}=\frac{1}{2}r^3_AH\bigg[-4H\left(\xi_0+\xi_1H+\xi_2\left(\frac{\dot{H}}{H}+H\right)\right)+\rho_m\bigg].
\end{equation}
The temperature of the apparent horizon can be obtained as \cite{setare mr}
\begin{equation}\label{57}
  T_h=\frac{1}{2\pi r_A}\left(1-\frac{\dot{r_A}}{2Hr_A}\right).
\end{equation}
Using equations \eqref{55}, \eqref{56} and \eqref{57}, we obtain
\begin{equation}\label{58}
  T_h\dot{S_h}=4\pi r^3_A H\bigg[\rho-4H\left(\xi_0+\xi_1H+\xi_2\left(\frac{\dot{H}}{H}+H\right)\right)\bigg]\bigg[1-\frac{\dot{r_A}}{2Hr_A}\bigg].
\end{equation}
By Gibbs equation, the change of entropy in the viscous fluid inside the apparent horizon can be defined as
\begin{equation}\label{59}
  T_mdS_m=d(\rho V)+\bar{p}dV
\end{equation}
where $T_m$ is the temperature of the bulk viscous fluid, $V=\frac{4}{3}\pi r^3_A$ is the volume. From the equations \eqref{15} and \eqref{33}, the Gibbs equation becomes
\begin{equation}\label{60}
  T_m dS_m=Vd\rho + \bigg[\rho-4H\left(\xi_0+\xi_1H+\xi_2\left(\frac{\dot{H}}{H}+H\right)\right)\bigg]dV.
\end{equation}
The temperature $T_m$ of the viscous fluid and the horizon $T_h$ are equal under equilibrium conditions. So, the Gibbs equation \eqref{60} becomes
\begin{eqnarray}\label{61}
  T_hS_m &=& 4\pi r^3_A H\bigg[4H\left(\xi_0+\xi_1H+\xi_2\left(\frac{\dot{H}}{H}+H\right)-\rho\right)\bigg]\nonumber \\
  &+&4\pi r^2_A \dot{r_A}\bigg[\rho-4H\left(\xi_0+\xi_1H+\xi_2\left(\frac{\dot{H}}{H}+H\right)\right)\bigg].
\end{eqnarray}
Adding equations \eqref{58} and \eqref{61}, yields
\begin{equation}\label{62}
  T_h(S_m+S_h)=\frac{A}{4}r^3_AH\bigg[\rho-4H\left(\xi_0+\xi_1H+\xi_2\left(\frac{\dot{H}}{H}+H\right)\right)\bigg]^2.
\end{equation}
Using the value of Hubble parameter and the energy density from the equations \eqref{27} and \eqref{45}, the equation \eqref{62} reduces to
\begin{eqnarray}\label{63}
  T_h(\dot{S_h}+\dot{S_m}) &=& \frac{(3n+1)Ar^3_A}{16}[(3n+1)t-2k]^{-1}\bigg[\frac{1}{8\pi+3\lambda-\gamma}\bigg[
  \xi_0\frac{(3n+1)(8\pi+4\lambda)}{(3n+1)t-2k} \nonumber \\
  &+&\frac{(\xi_1-3\xi_2)(3n+1)^2-12n(n+1)}{4[(3n+1)t-2k]^2}\bigg] \nonumber \\
  &-&\frac{3n+1}{4}[(3n+1)t-2k]^{-1}\left(\xi_0+(\xi_1-3\xi_2)\frac{3n+1}{4}\frac{1}{(3n+1)t-2k}\right)\bigg]^2
\end{eqnarray}
From the equation \eqref{63}, we observe that $\dot{S_h}+\dot{S_m} \ge 0$ for all $'t'$. This implies that the generalized second law of thermodynamics is valid throughout the evolution.

\section{Conclusions}
The usual Kaluza-Klein philosophy is to assume that the radius of the compactified circle is very, very small (by small, it means that radius is roughly, speaking, of the order the Planck length, $10^{-33}$ centimeters). As per the Kaluza-Klein dimensional reduction process, the three standard spatial dimension will be expanded and the extra dimension must reduce to a Planckian length. We have to find the range of $n$, so that our models should obey the Kaluza-Klein dimensional reduction process.
\begin{itemize}
  \item For $n<\frac{-1}{3}$, the scale factor $b$ of extra dimension decreases and the scale factor $a$ increases as $t\rightarrow\infty$, i. e. the size of the extra dimension reduces to a Planckian length (unobservable length) and the size of three standard spatial dimension expand as $t\rightarrow\infty$.
   \item When $t\rightarrow\infty$, $b\rightarrow 0$, hence the fifth dimension is ruled out at $t\rightarrow\infty$. Thus, there is a less significance of Kaluza-Klein theory at infinite time than the present.
\end{itemize}
From section-4, the cosmological model discussed with perfect fluid. We found that, the time dependent Hubble parameter and the time dependent energy density, but the deceleration parameter is constant $(\frac{-1}{2})$. The following observations are made from section-4:
\begin{itemize}
  \item The Hubble parameter is always positive and reduces as time increases, finally  $H\rightarrow\ 0$ as $t\rightarrow\infty$. Hence, we found that $H>0$ and $q<0$, therefore our model is expanding and accelerating.
   \item For $\gamma=0$, the energy density from \eqref{28} becomes constant, which leads to a false vacuum case.
  \item For $n<\frac{-1}{3}$ and $\gamma=2, \frac{2}{3}$, the Null Energy Condition (NEC) is satisfied $(p+\rho\ge0)$, the Weak Energy Condition (WEC) is satisfied ($\rho\ge0$ and $p+\rho\ge0$), i. e. the positive of the energy density for any observer at any point and also, the Strong Energy Condition (SEC) is satisfied ($p+\rho\ge0$ and $\rho+3p\ge0$).
  \item For $\gamma=1$, the energy density $\rho$ becomes negative for any finite time '$t$', hence the model violet WEC and SEC where as the NEC is satisfied. The energy density can be positive for $n>\frac{-1}{3}$, but which leads to contradict the Kaluza-Klein dimensional reduction process. Therefore, the model is not compatible for $\gamma=1$.
   \item For $\gamma=\frac{1}{2}$, the energy density $(\rho)$ becomes imaginary with irrespective of time $'t'$, in general, for $\gamma=\frac{1}{2}\times \frac{l}{m}$, where $0\le\frac{l}{m}\le4$, $l>0, m>0$ and $l$ is not an even number, the energy density $(\rho)$ becomes imaginary with irrespective of time $'t'$. This is not physically realistic. Hence the solutions of the model is not acceptable for $\gamma=\frac{1}{2}\times \frac{l}{m}$, i. e. the model is not compatible.
   \item In general, for $\gamma=2\times\frac{l}{m}$, where $0\le \frac{l}{m}\le 1$, $l>0, m>0$ and $m$ is not an even number, the model satisfies all energy conditions. Hence, we found, one acceptable model for $\gamma=2\times\frac{l}{m}$.
   \item $a\rightarrow a_s$, $\rho\rightarrow\infty$ and $|p|\rightarrow\infty$, as $t\rightarrow t_s$, hence type-III singularity is observed. Also, we found that the model posses type-I singularity for $n<\frac{-1}{3}$, because $a\rightarrow a_s$, $\rho\rightarrow\infty$, $|p|\rightarrow\infty$ as $t\rightarrow t_s$ for $n<\frac{-1}{3}$.

\end{itemize}
The following observations are made from section-5 (cosmology with constant bulk viscous coefficient $(\xi=\xi_0)$):
\begin{itemize}
\item From the equations \eqref{28} and \eqref{32}, we observed that the energy density with constant bulk viscous coefficient is dominated by the energy density with no viscous fluid.
  \item For $\gamma=1$, the energy density from \eqref{32} becomes infinity irrespective of time, i. e. $a\rightarrow a_s$, $\rho\rightarrow\infty$, $p=0$ as $t\rightarrow t_s$. Hence the model either posses a different kind of singularity or the model is incompatible.
  \item For $\gamma=0$, the energy density becomes negative. It violates WEC where as it satisfies NEC and SEC.
  \item For $\gamma=2$, the behavior of the energy density, energy conditions and the behavior of the model are exactly same as the case model with no viscous fluid.
  \item If $\gamma=\frac{2}{3}$, then $\rho<0$, $p+\rho \le 0$ and $\rho+3p \ge 0 $, it violet all energy conditions, whereas the model with no viscosity satisfied all energy conditions for $\gamma=\frac{2}{3}$.
  \item In general, the model does not satisfy any energy conditions for $\gamma=2\times\frac{l}{m}$, where $0\le \frac{l}{m}< \frac{1}{2}$, $l>0, m>0$ and $m$ is not an even number, whereas the model satisfies all energy conditions for $\gamma=2\times\frac{l}{m}$, where $\frac{1}{2}\le \frac{l}{m}\le 1$, $l>0, m>0$ and $m$ is not an even number.
  \item For $\gamma=\frac{1}{2}$, the energy density $(\rho)$ becomes imaginary with irrespective of time $'t'$, in general, for $\gamma=\frac{1}{2}\times \frac{l}{m}$, where $0\le\frac{l}{m}\le4$, $l>0, m>0$ and $l$ is not an even number, the energy density $(\rho)$ becomes imaginary with irrespective of time $'t'$. This is not physically realistic. Hence the solutions of the model is not acceptable for $\gamma=\frac{1}{2}\times \frac{l}{m}$, i. e. the model is not compatible.
\end{itemize}
The following observations are made from section-6 (cosmology with non constant bulk viscous coefficient):
\begin{itemize}
  \item The behavior of the energy density from the equations \eqref{35}, \eqref{38} and \eqref{41} are approximately same.
  \item For $\gamma=2$, the energy density from the equations \eqref{35}, \eqref{38} and \eqref{41} are become infinity irrespective of time, i. e. $a\rightarrow a_s$, $\rho\rightarrow\infty$, $p\rightarrow\infty$ as $t\rightarrow t_s$. Hence the model posses type-III singularity.
  \item For $\gamma=0$, the energy density from the equations \eqref{35}, \eqref{38} and \eqref{41} are positive throughout the evolution and, we observed that the NEC and WEC are valid whereas the SEC is violated. The violation of SEC represents the accelerated expansion of the universe.
  \item For $\gamma=1$, the second term is dominated by the first term in each equation \eqref{35}, \eqref{38} and \eqref{41}. Therefore, the energy density is positive for $k_4<0, k_5<0$ and $k_6<0$, and satisfies all energy conditions.
   \item For $\gamma=\frac{4}{3}, \frac{2}{3}$, the energy density becomes positive and satisfies all energy conditions, when $k_4>0, k_5>0$ and $k_6>0$.
   \item In general the model satisfies all the energy conditions (NEC, WEC \& SEC) for $\gamma=2\times\frac{l}{m}$, where $\frac{1}{3}\le \frac{l}{m}\le 1$, $l>0, m>0$ and $m$ is not an even number, whereas the model violates SEC for $\gamma=2\times\frac{l}{m}$, where $0\le \frac{l}{m}< \frac{1}{3}$, $l>0, m>0$ and $m$ is not an even number.

   \item From the equation \eqref{43}, it is observed that the energy density is always positive for all time $'t'$. The model satisfies all the energy conditions for $\frac{2}{3}\le \gamma\le 2$, whereas the model violates the SEC for $0\le\gamma<\frac{2}{3}$.
\end{itemize}
In this paper, the authors discussed the bulk viscous fluid Kaluza-Klein cosmological models with the validity of the second law of thermodynamics and the generalized second law of thermodynamics in $f(R, T)$ gravity. We observed that, the second law of thermodynamics is violated for Kaluza-Klein bulk viscous fluid $f(R, T)$ gravity model, when the bulk viscous coefficient $\xi$ is proportional to $\frac{\ddot{a}}{\dot{a}}$ (i. e. $\xi=\xi_2\frac{\ddot{a}}{\dot{a}}$). However, the generalized second law of thermodynamics is valid throughout the evolution. Otherwise, the second law of thermodynamics is valid for other models, say model described by ($\xi=\xi_1\frac{\dot{a}}{a}$, $\xi=\xi_1\frac{\dot{a}}{a}+\xi_2\frac{\ddot{a}}{\dot{a}}$ and $\xi=\xi_0+\xi_1\frac{\dot{a}}{a}+\xi_2\frac{\ddot{a}}{\dot{a}}$). The model described by the bulk viscous coefficient $\xi=\xi_0+\xi_1\frac{\dot{a}}{a}+\xi_2\frac{\ddot{a}}{\dot{a}}$ does not contain singularity, whereas the model described by the bulk viscous coefficient ($\xi=\xi_0, \xi=\xi_1\frac{\dot{a}}{a}$ and $\xi=\xi_1\frac{\dot{a}}{a}+\xi_2\frac{\ddot{a}}{\dot{a}}$) contains singularity. Hence, from the above discussions and observations, we may conclude that the model described by the bulk viscous fluid, where the bulk viscous coefficient $\xi=\xi_0+\xi_1\frac{\dot{a}}{a}+\xi_2\frac{\ddot{a}}{\dot{a}}$ is more realistic than the others (i. e. $\xi=\xi_0, \xi=\xi_1\frac{\dot{a}}{a}$ and $\xi=\xi_1\frac{\dot{a}}{a}+\xi_2\frac{\ddot{a}}{\dot{a}}$). Finally, from the statefinder parameter, we observed that $r<1$ and $s<1$, which is different from the $\wedge$CDM model.


\begin{figure}[ht!]
  \centering
  \includegraphics[width=10cm,height=8cm]{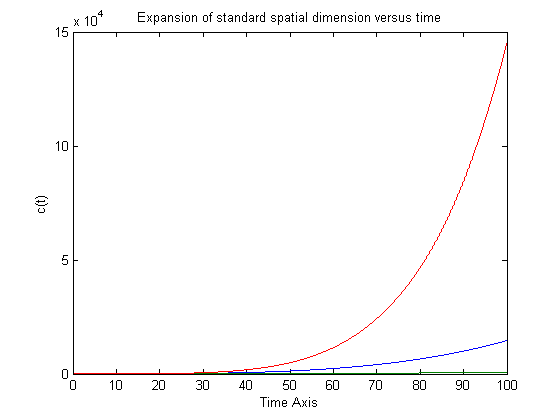}
 \caption{From the figure-1, we observed that the standard spatial dimension is expanding for large $t$, when $n<\frac{-1}{3}$. }\label{1}
\end{figure}
\begin{figure}[ht!]
  \centering
  \includegraphics[width=10cm,height=8cm]{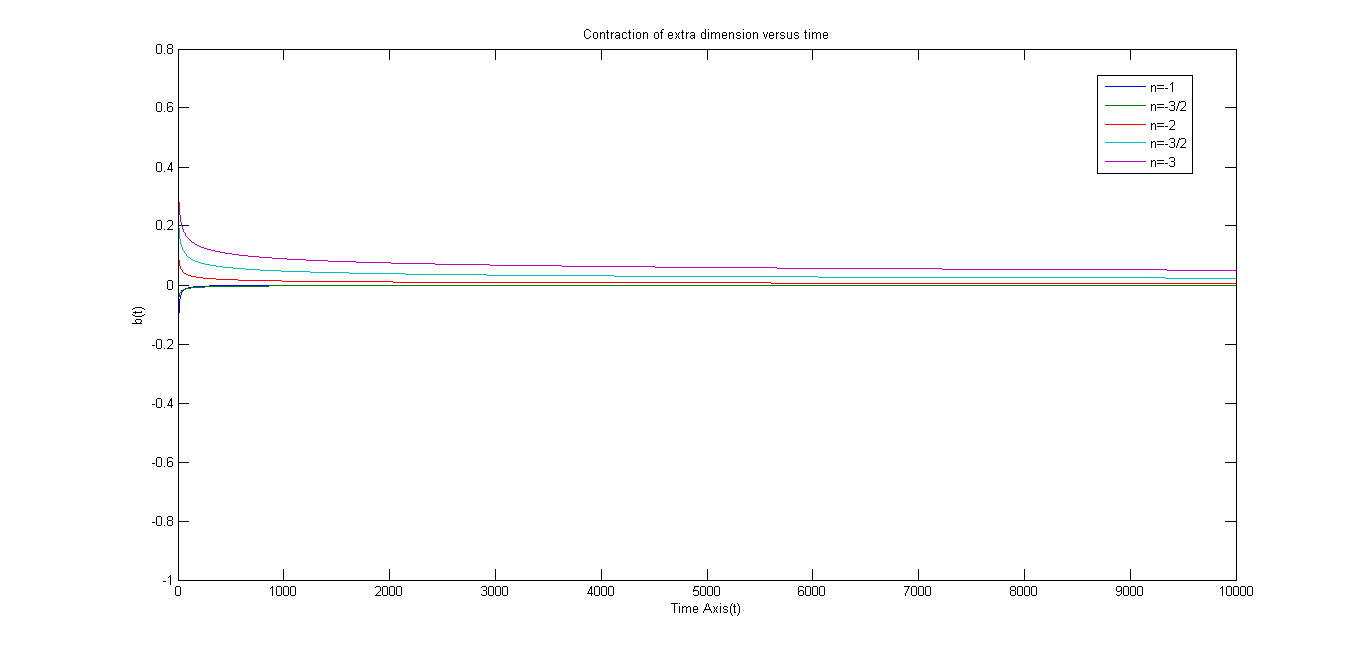}
 \caption{This figure indicates that the scale factor $b$ of extra dimension reduces to a Planckian length (unobservable length) as $t\rightarrow\infty$ for $n<\frac{-1}{3}$. The process of this contraction of extra dimension is called the Kaluza-Klein dimensional reduction process or Kaluza-Klein compactification of extra dimension.}\label{2}
\end{figure}
\newpage
\begin{figure}[ht!]
  \centering
  \includegraphics[width=10cm,height=8cm]{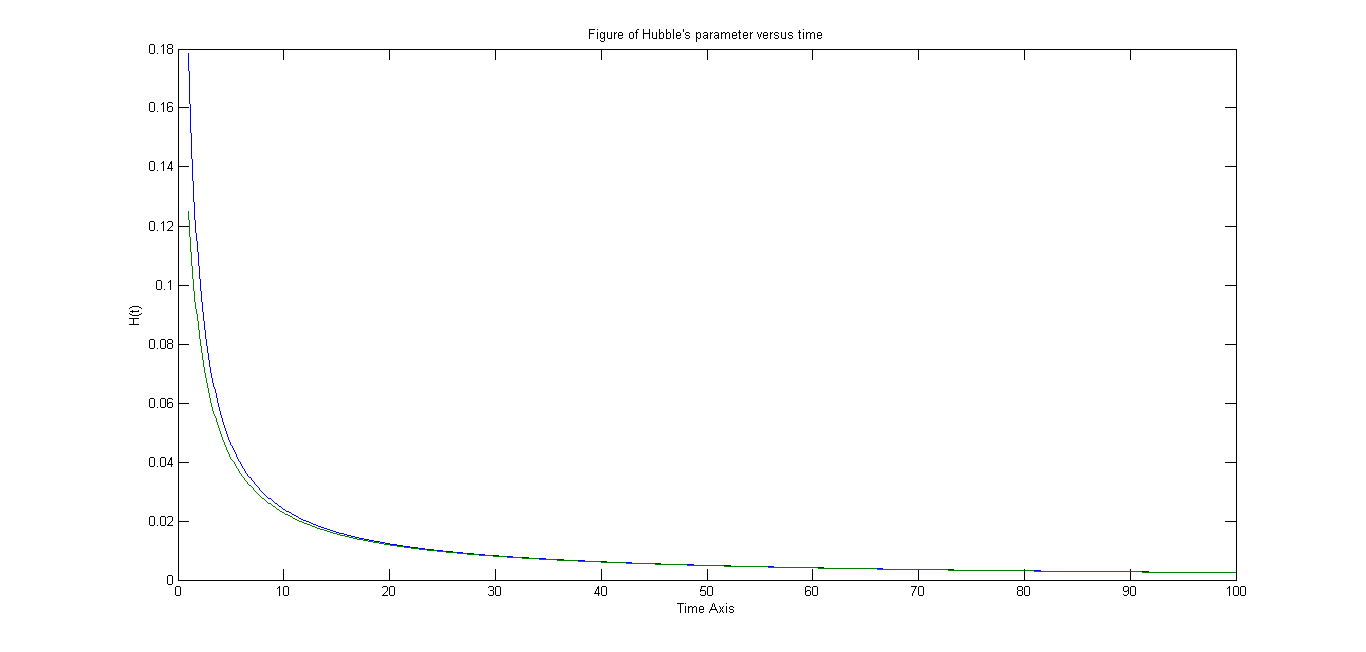}
 \caption{This figure indicates that the value of the Hubble parameter is always positive and decreases to zero as $t\rightarrow\infty$. Therefore, the universe is in expanding in nature.}\label{3}
\end{figure}

\begin{figure}[ht!]
  \centering
  \includegraphics[width=10cm,height=8cm]{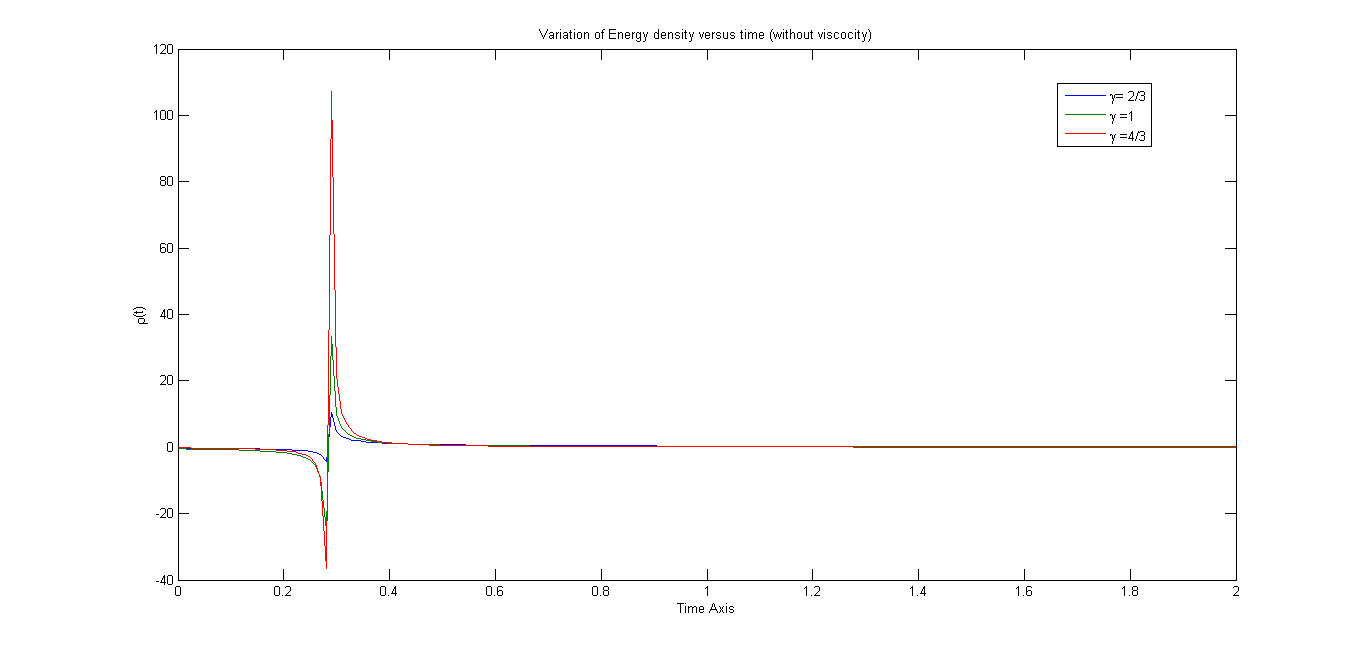}
 \caption{This figure indicates that the variation of the energy density with respect to time when the source of gravitation is a perfect fluid i. e. no viscosity. This shows that the model posses a type-III singularity.}\label{4}
\end{figure}
\newpage
\begin{figure}[ht!]
  \centering
  \includegraphics[width=10cm,height=8cm]{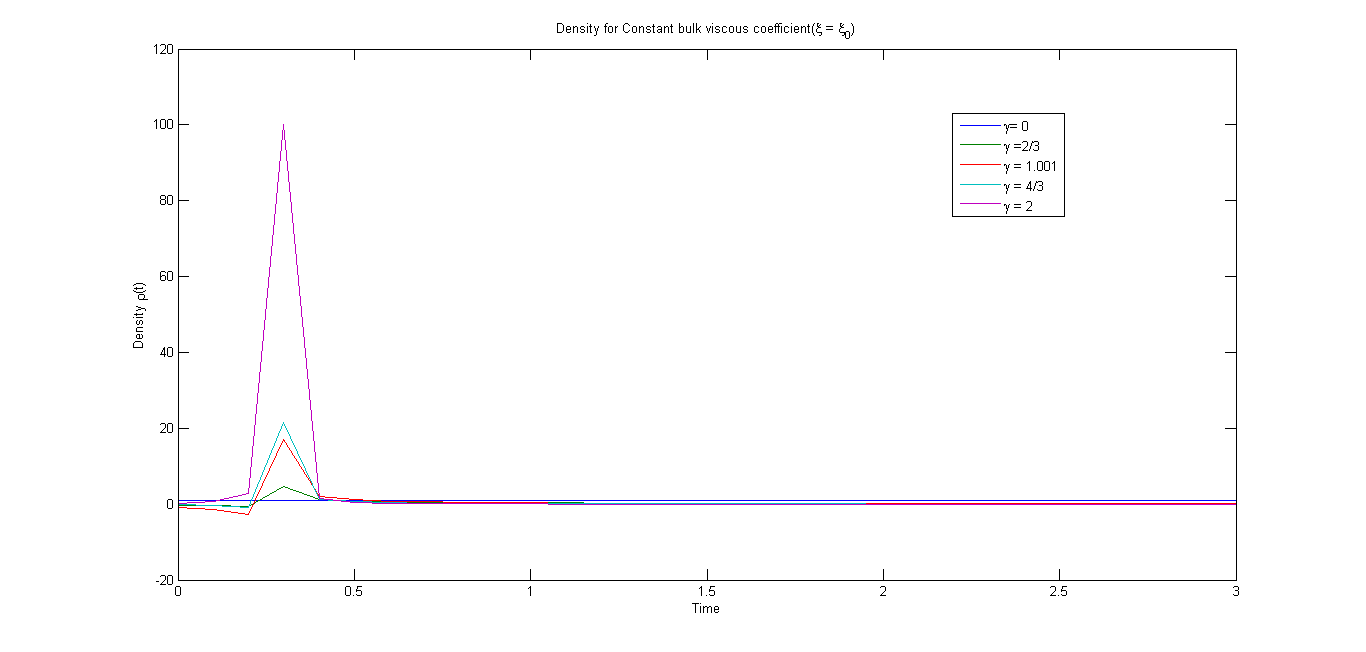}
 \caption{This figure indicates that the variation of the energy density with respect to time when the source of gravitation is a combination of perfect fluid and bulk viscous fluid, where the bulk viscous coefficient $\xi$ is simply constant, i. e. $\xi=\xi_0$. This shows that the model possess a singularity for the dust case $(\gamma=1, i. e. p=0)$.}\label{5}
\end{figure}
\begin{figure}[ht!]
  \centering
  \includegraphics[width=10cm,height=8cm]{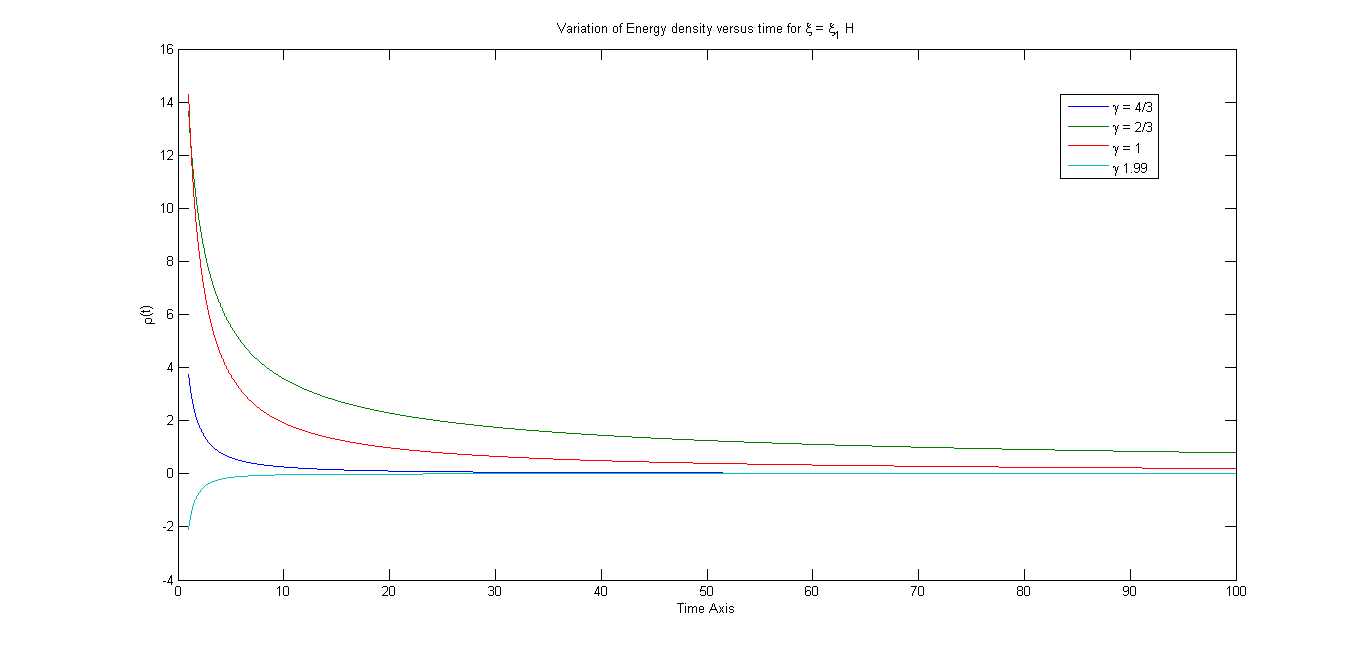}
 \caption{This figure indicates that the variation of the energy density with respect to time when the source of gravitation is a combination of perfect fluid and bulk viscous fluid, where the bulk viscous coefficient is proportional to the expansion rate of the universe, i. e. $\xi=\xi_1\frac{\dot{a}}{a}$. This shows that the model possess a type-III singularity for stiff fluid case $(\gamma=2, i. e. p=\rho)$.}\label{6}
\end{figure}
\newpage
\begin{figure}[ht!]
  \centering
  \includegraphics[width=10cm,height=8cm]{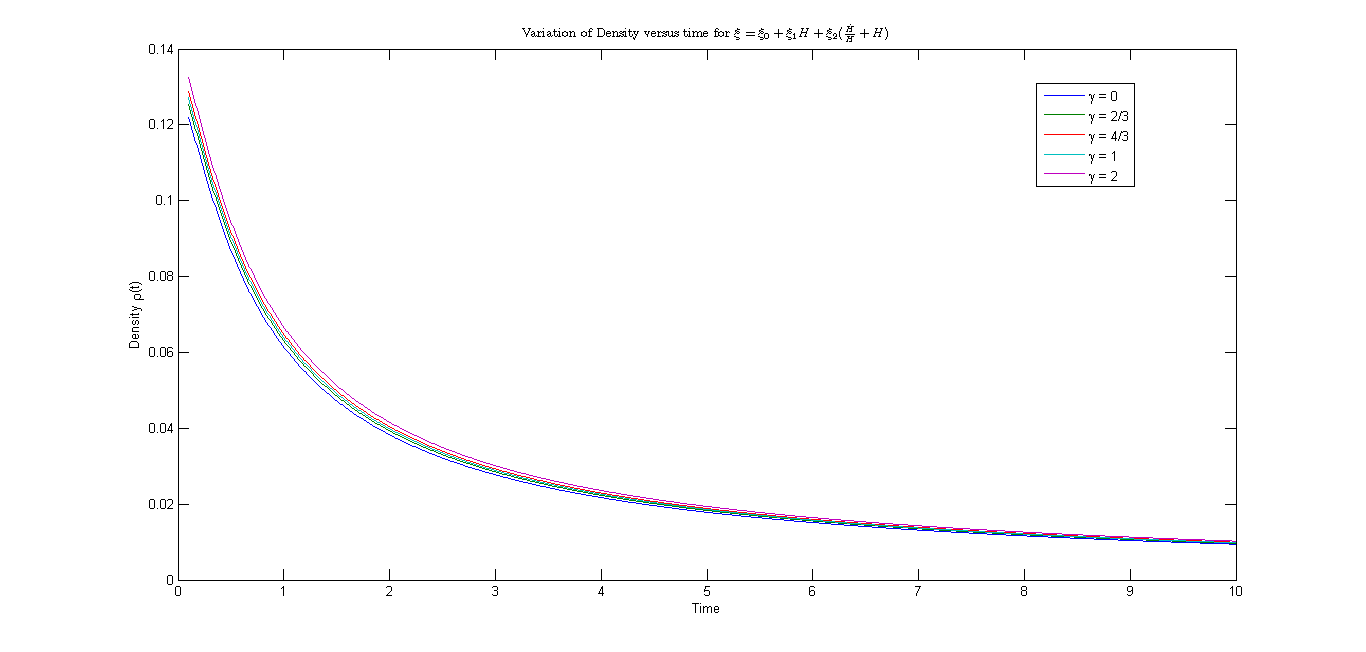}
 \caption{This figure indicates that the variation of the energy density with respect to time when the source of gravitation is a combination of perfect fluid and bulk viscous fluid, where the bulk viscous coefficient is proportional to the linear combination of expansion rate of the universe and acceleration of the expansion of the bulk viscosity, i. e. $\xi=\xi_0+\xi_1\frac{\dot{a}}{a}+\xi_2\frac{\ddot{a}}{\dot{a}}$. }\label{7}
\end{figure}
\begin{figure}[ht!]
  \centering
  \includegraphics[width=10cm,height=8cm]{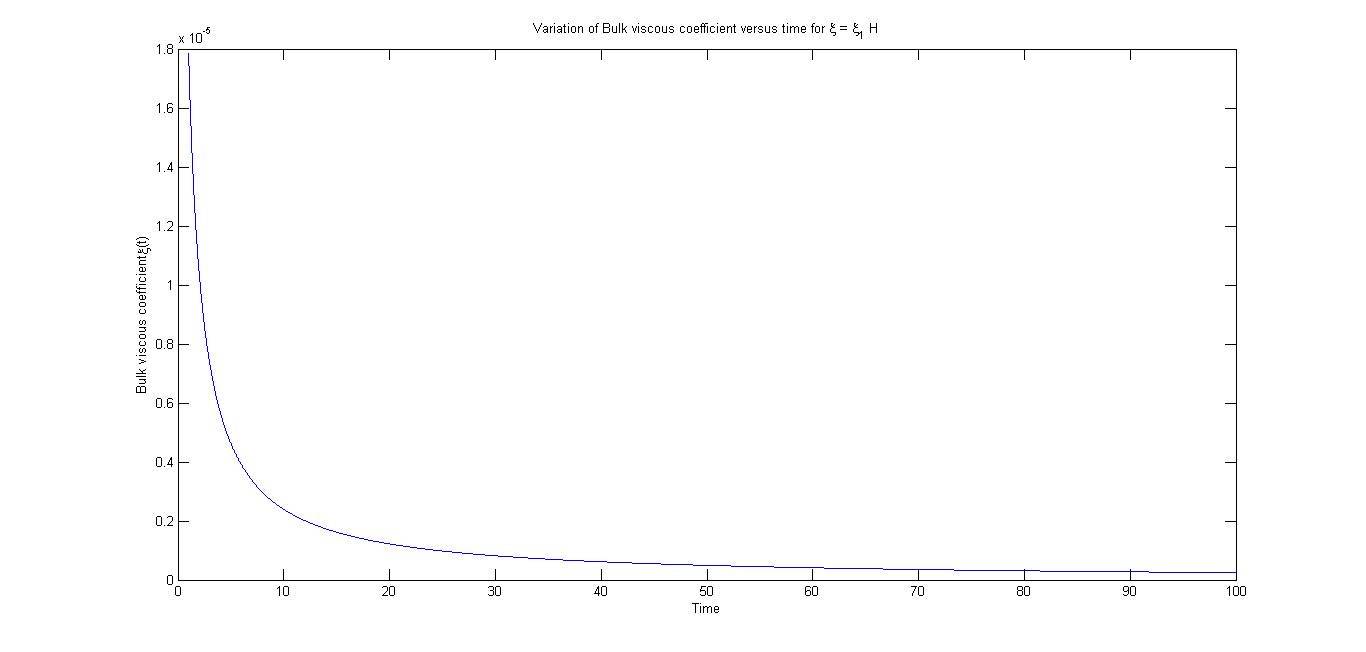}
 \caption{This figure indicates that the variation of the bulk viscous coefficient $(\xi)$ with respect to time, when $(\xi)$ is proportional to the expansion rate of the universe (i. e. $\xi=\xi_1\frac{\dot{a}}{a})$. This shows that $(\xi)$ is always positive and decreases to zero as $t\rightarrow\infty.$}\label{8}
\end{figure}
\newpage

\begin{figure}[ht!]
  \centering
  \includegraphics[width=10cm,height=8cm]{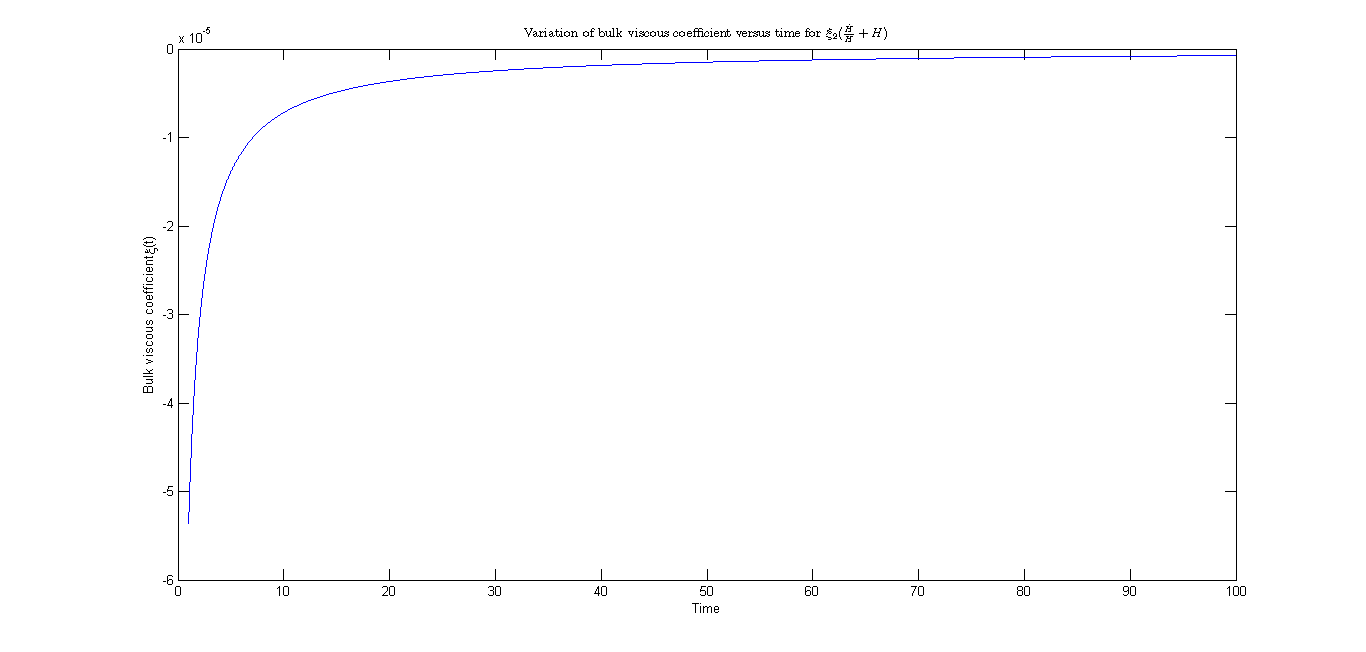}
 \caption{This figure indicates that the variation of the bulk viscous coefficient $(\xi)$ with respect to time, when $(\xi)$ is proportional to the acceleration of the expansion of universe (i. e. $\xi=\xi_2\frac{\ddot{a}}{\dot{a}})$. This shows that $(\xi)$ is always negative and decreases to zero as $t\rightarrow\infty$, which violate the second law of thermodynamics.}\label{9}
\end{figure}

\begin{figure}[ht!]
  \centering
  \includegraphics[width=10cm,height=8cm]{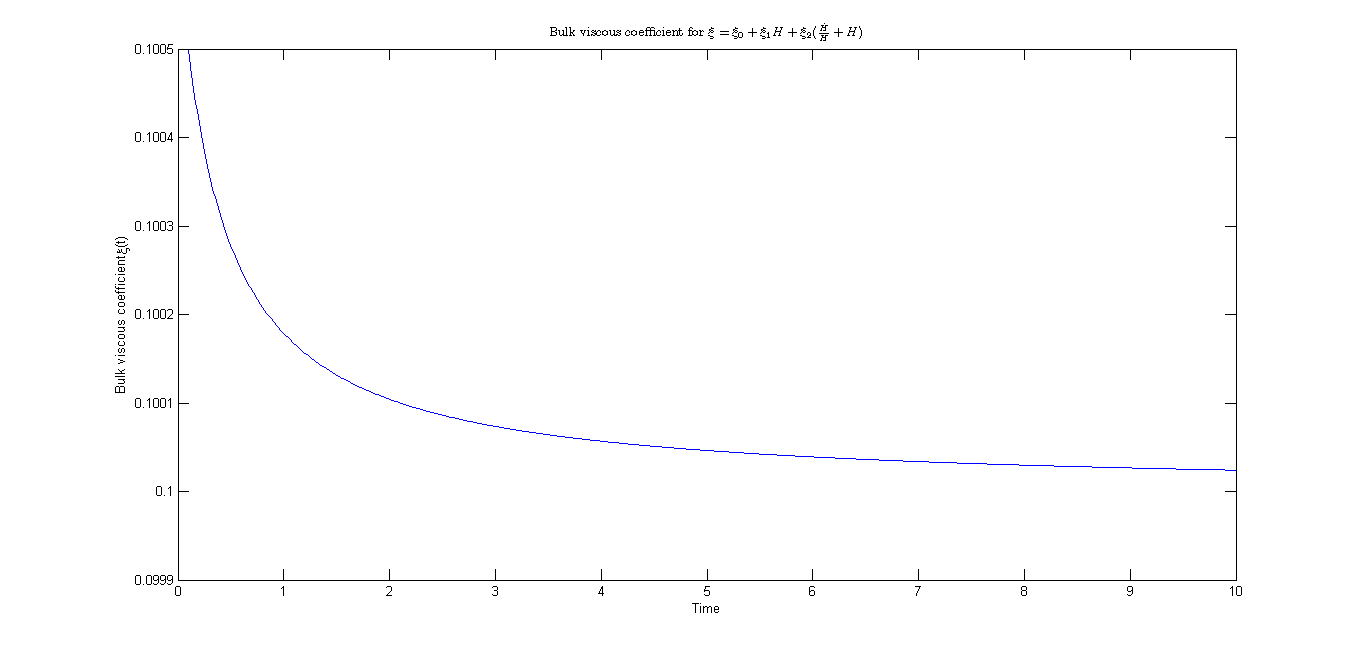}
 \caption{This figure indicates that the variation of the bulk viscous coefficient $(\xi)$ with respect to time, when $(\xi)$ is proportional to the linear combination of expansion rate and the acceleration of the universe (i. e. $\xi=\xi_0+\xi_1\frac{\dot{a}}{a}+\xi_2\frac{\ddot{a}}{\dot{a}})$, where $\xi_0, \xi_1$ and $\xi_2$ are positive constants. This shows that $(\xi)$ is always positive and decreases to $\xi_0$ as $t\rightarrow\infty.$}\label{10}
\end{figure}

\end{document}